\documentclass[useAMS,usenatbib,usegraphicx,letterpaper]{mn2e}

\usepackage{multirow} 
\usepackage{epsfig}
\usepackage{subfigure}

\newcommand{\etal}{et al.}

\def\mnras{MNRAS}
\def\apj{ApJ}
\def\apjs{ApJS}
\def\aap{A\&A}
\def\aj{AJ}

\title[PHL~1092: a transient X--ray weak QSO] {PHL~1092 as a
  transient extreme X--ray weak quasar} \author[G.\ Miniutti \etal]
      {G. Miniutti$^{1}$\thanks{gminiutti@laeff.inta.es},
        A.~C. Fabian$^2$, W.~N. Brandt$^3$, L.~C. Gallo$^4$ and
        Th. Boller$^5$
\\ \\ 
  $^1$ LAEX, Centro de Astrobiologia (CSIC--INTA); LAEFF, P.O: Box 78, E-28691, Villanueva de la Ca\~nada, Madrid, Spain\\
  $^2$Institute of Astronomy, Madingley Road, Cambridge CB3 0HA
  \\
  $^3$ Department of Astronomy and Astrophysics, The Pennsylvania
  State University, 525 Davey Lab., Univeristy Park, PA 16802, USA\\
  $^4$ Department of Astronomy \& Physics, Saint Mary's University, 923 Robie Street, Halifax, NS B3H 3C3 \\
  $^5$ Max--Planck--Institut f\"ur extraterrestrische Physik, Postfach
  1312, 85741 Garching, Germany
}

\pagerange{\pageref{firstpage}--\pageref{lastpage}}
\pubyear{2001}

\usepackage{times}

\begin{document}

\label{firstpage}

 \maketitle

\begin{abstract}
We report a dramatic variability event in the X--ray history of the
Narrow--Line quasar PHL~1092 ($z=0.396$). Our latest 2008 {\it
  XMM--Newton} observation reveals a flux drop of $\sim$~200 with
respect to the previous observation performed about 4.5 years earlier,
and a drop of $\sim$~135 with respect to its historical flux. Despite
the huge X--ray variation, the UV flux remains constant producing a
very significant steepening of the optical to X--ray slope
$\alpha_{\rm{ox}}$ from $-$1.56 to $-$2.44, making PHL~1092 one of the
most extreme X--ray weak quasars. The similarity in the soft X--ray
spectral shape between the present and previous observations, together
with the persistent UV flux and the lack of any dramatic change in the
optical spectrum suggest that an absorption event is not likely to be
the origin of the observed variation. If absorption is ruled out, the
sudden X--ray weakness of PHL~1092 must be produced by a transient
significant weakening or disruption of the X--ray emitting corona.
\end{abstract}


\section{Introduction}

PHL~1092 ($z=0.396$) is a radio--quiet quasar (QSO) characterized in
the optical band by outstanding Fe~\textsc{ii} emission with a
Fe~\textsc{ii}$\lambda$4570~/~H$\beta$ ratio of $\sim$~5.3 (Bergeron
\&Kunth 1980; Bergeron \&Kunth 1984Kwan et al 1995). Its line widths
of $\sim$~1800~km~s$^{-1}$ and
[O~\textsc{iii}]$\lambda$5007~/~H$\beta$ ratio of $\sim$~0.9
classify PHL~1092 as a high--luminosity representative of the
Narrow--Line Seyfert~1 (NLS1) galaxy population (Osterbrock \& Pogge
1985; Goodrich 1989). In fact, strong Fe~\textsc{ii} emission seems to
be a further characteristics of the optical spectra of NLS1 galaxies
(e.g. V{\'e}ron-Cetty, V{\'e}ron \& Gon{\c c}alves 2001).  NLS1
galaxies also exhibit remarkable properties in the X--ray regime.
They are characterized by large--amplitude short--timescale X--ray
variability and their X--ray spectra are steeper than in Broad--Line
sources (e.g. Boller, Brandt \& Fink 1996; Brandt, Mathur \& Elvis
1997; Leighly 1999a,b). This is more dramatic in the soft X--ray band,
where a prominent soft excess of debated origin is very often
observed. A number of NLS1 are X--ray weak if compared to
optically--selected standard QSO and their X--ray variability is
generally larger than in the optical/UV producing changes in the
optical to X--ray spectral slope $\alpha_{\rm{ox}}$ up to $\Delta
\alpha_{\rm{ox}} \sim 0.3-0.4$ (e.g. NGC~4051, 1H~0707--495, see Gallo
2006).

PHL~1092 is one of the most X--ray variable NLS1 known, despite its
QSO--like X--ray and optical luminosities (see also the remarkable
case of IRAS~13224--3809, Boller et al. 1997). During the 1997 ROSAT
monitoring of PHL~1092, a maximum variability amplitude of a factor
$\sim$~14 was reported (Brandt et al. 1999). The typical luminosity in
the 0.2-2~keV band is $\sim 1\times 10^{45}$~erg~s$^{-1}$. 

PHL~1092 was observed with {\it XMM--Newton} on three occasions. The
first observation was performed on 2000 July 31 and lasted
$\sim$~32~ks. The ODF files for the EPIC cameras could not be
recovered, but X--ray and UV light curves from the RGS and Optical
Monitor (OM) could be obtained (Gallo et al. 2004). The source was
then re--observed three years later on 2003 July 18 for $\sim$~29~ks,
resulting in complete set of ODF files for all detectors. Finally, we
obtained a further $\sim$~60~ks observation on 2008 January 20 which
is the main subject of this Letter. Besides the simultaneous UV
observations with the OM, we also obtained quasi--simultaneous optical
spectra at the Hobby--Eberly Telescope (HET, 600~s on 2008 January 30)
and at the William Herschel telescope (WHT, 900~s on 2008 February
14).

\section{Dramatic long--term X--ray variability}

\begin{figure}
\begin{center}
\includegraphics[width=0.44\textwidth,height=0.28\textwidth,angle=0]{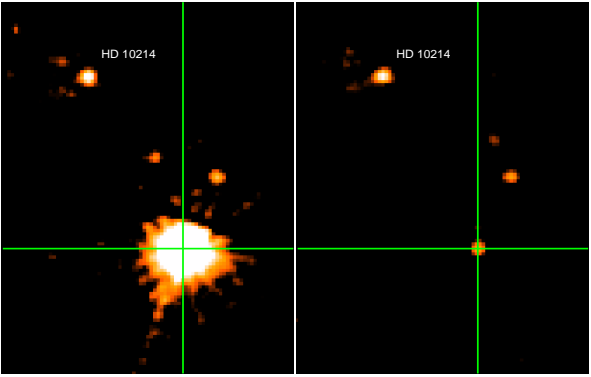}
\caption{The 0.2--0.9~keV EPIC pn images of the PHL~1092 region from the 2003
  (left) and 2008 (right) observations. Astrometry has been checked by
  cross--correlating the X--ray images with the USNO A~2.0
  catalogue. One of the common sources (the G5 star HD~10214) is
  labelled. Cross--hairs are locked at the nominal coordinates of
  PHL~1092 and the images share the same scale (about 5$\times$7
  arcmin.) and color--bar scheme. The raw images have been slightly
  smoothed for clarity.}
\label{images}
\end{center}
\end{figure}

\begin{figure}
\begin{center}
\includegraphics[width=0.28\textwidth,height=0.42\textwidth,angle=-90]{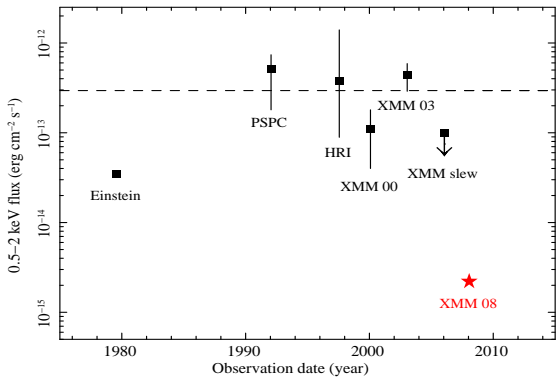}
\caption{The historical 0.5--2~keV flux of PHL~1092. Count rates have
  been converted in fluxes using {\small{PIMMS}}. The vertical lines
  display the intra--observation flux variability range (when
  available). From earlier to latest, fluxes from {\it{Einstein}}/IPC
  (Wilkes et al. 1994), {\it{ROSAT}}/PSCP (Forster \& Halpern 1996),
  {\it{ROSAT}}/HRI (Brandt et al. 1999), {\it XMM--Newton} (2000 and
  2003, Gallo et al. 2004), a {\it{XMM--Newton}} slew passage, and our
  latest 2008 {\it XMM--Newton} observation are shown. A further {\it
    ASCA} observation, simultaneous with the ROSAT HRI monitoring
  (Leighly et al. 1999a,b) is not reported here since flux and
  variability are encompassed by the HRI data point. The horizontal
  line shows the average flux excluding the {\it XMM--Newton} slew
  upper limit and the last observation ($\sim 3\times
  10^{-13}$~erg~cm$^{-2}$~s$^{-1}$).}
\label{lchist}
\end{center}
\end{figure}

The most striking result of our new 2008 {\it XMM--Newton} observation
is that PHL~1092 is much fainter than during any previous X--ray
observation. This is readily visible by examining X--ray images of the
region. In Fig.~\ref{images} we show the EPIC pn images from the two
available good quality {\it XMM--Newton} observations clearly showing
that a dramatic variability event has taken place, so that PHL~1092 is
not even the brightest X--ray source in the region in 2008. The
0.5--2~keV flux of PHL~1092 from all previous available X--ray
observations is shown in Fig.~\ref{lchist}. As will be discussed
below, the flux drop between the latest {\it XMM--newton} pointed
observation in 2003 is about a factor 200, while a drop of a factor
$\sim$135 is observed with respect to the average historical flux
($\sim 3\times 10^{-13}$~erg~cm$^{-2}$~s$^{-1}$).

\begin{figure}
\begin{center}
\includegraphics[width=0.28\textwidth,height=0.42\textwidth,angle=-90]{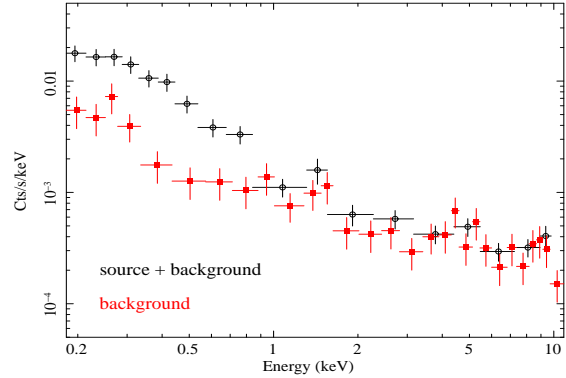}
\caption{The source plus background spectrum of PHL~1092 from the 2008
  {\it XMM--Newton} observation is compared with the background from a
  nearby region.}
\label{srcback}
\end{center}
\end{figure}

In Fig.~\ref{srcback}, we compare the 2008 EPIC pn source plus
background spectrum of PHL~1092 with the background itself. PHL~1092
is well detected above the background level up to $\sim$~0.9~keV
only. Hence we here focus on the 0.2-0.9~keV range only, corresponding
to the $\sim$0.28--1.26~keV band in the rest--frame. Given that the
main interest of our work is to assess the long--term X--ray
variability of PHL~1092, we proceed by analysing the new 2008 data
together with the earlier 2003 {\it XMM--Newton} observation, which
represents a typical X--ray flux state of the source (see
Fig.~\ref{lchist}.)

\subsection{Comparison with the 2003 {\it XMM--Newton} observation}

\begin{figure}
\begin{center}
\includegraphics[width=0.28\textwidth,height=0.42\textwidth,angle=-90]{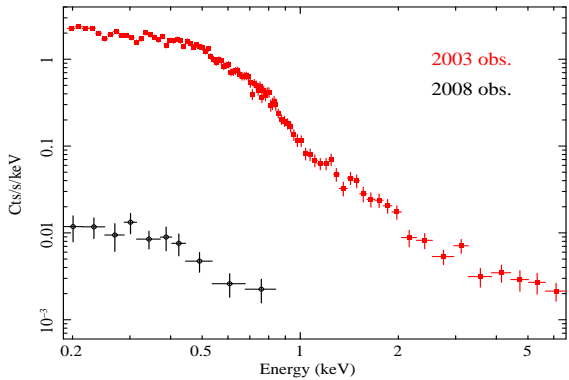}
\caption{A comparison between the 2003 and 2008 {\it XMM--Newton} EPIC
  pn spectra of PHL~1092. The huge flux drop is clearly visible in the
  soft band, where data are available for both pointings.  }
\label{speccomp}
\end{center}
\end{figure}

In Fig.~\ref{speccomp} we show the EPIC--pn spectrum from the 2008
observation together with the 2003 data (above $\sim$7~keV the
background dominates during the 2003 observation). In order to compare
the source fluxes during the 2003 and 2008 {\it XMM--Newton}
observations in more detail, we apply the same spectral model to the
two data sets, and we first adopt a simple power law model absorbed by
a column of neutral gas which we allow to vary in the range of the
measured Galactic value ($3.6\pm 0.2\times 10^{20}$~cm$^{-2}$, Murphy
et al. 1996)

The 2008 data are well described ($\chi^2=12$ for 23 degrees of
freedom, dof) by a very steep but loosely constrained photon index
$\Gamma= 4.5\pm 0.7$ . We measure a flux of $(7.0\pm 3.5)\times
10^{-15}$~erg~cm$^{-2}$~s$^{-1}$ in the 0.2--0.9~keV band,
corresponding to a 0.2--0.9~keV unabsorbed luminosity of $\sim
3.1\times 10^{43}$~erg~s$^{-1}$. The 2003 data are
only relatively well described ($\chi^2=162$ for 130 dof), but we can
nonetheless obtain a good measure of the spectral shape with
$\Gamma=4.07\pm 0.05$ with a 0.2--0.9~keV flux of $1.28\pm 0.06\times
10^{-12}$~erg~cm$^{-2}$~s$^{-1}$, corresponding to an unabsorbed
luminosity of $\sim 4.1 \times 10^{45}$~erg~s$^{-1}$ in the same
energy range. 

Given that the 0.2--0.9~keV spectral shape is consistent with being
the same in the 2003 and 2008 observations, we then consider a joint
fit to the two data sets (in the 0.2--4~keV range for the 2003
observation to avoid the high--energy complexities discussed by Gallo
et al 2004, and in the 0.2--0.9~keV range for the 2008 one). We use
the simplest best--fitting model for the higher flux 2003 spectrum,
comprising a phenomenological disc blackbody (BB) model plus a power
law, modified by neutral absorption fixing all parameters to be the
same except for an overall normalization. We obtain an acceptable
joint fit ($\chi^2_\nu = 1.2$ for 202 dof) with kT$\sim$~100~eV,
$\Gamma \sim 2.6$ and N$_{\rm{H}} = 2.6 \pm 0.3 \times
10^{20}$~cm$^{-2}$ (less than the Galactic value). If the column
density is allowed to vary between the two observations, no
significant improvement is obtained and the 2008 data are inconsistent
with N$_{\rm H} \geq 6\times 10^{20}$~cm$^{-2}$. The BB temperatures
are also consistent with each other. By using the common spectral
model, we infer that the flux drop during the 2008 pointing with
respect to the previous one is of a factor $200\pm 40$. If the low
flux state is compared with the historical one shown in
Fig.~\ref{lchist}, the drop is of a factor $\sim 135$. Incidentally,
if the two flux states are indeed associated with the same spectral
component in the soft band, our result implies that interpreting the
soft X--ray excess in NLS1 galaxies as BB emission is implausible as
the luminosity drop occurs here at fixed temperature without following
the standard BB--defining $L_{\rm{BB}} \propto T^4$ relationship. As
shown in Fig.~\ref{srcback}, the hard X--rays during the 2008
observation are background--dominated. A simple power law fit to the
2--10~keV data is however useful to set a 2--10 flux upper limit of
$\leq 9.9\times 10^{-15}$~erg~s$^{-1}$~cm$^{-2}$ for future
reference. 

\subsection{X--ray variability}

\begin{figure}
\begin{center}
\includegraphics[width=0.28\textwidth,height=0.42\textwidth,angle=-90]{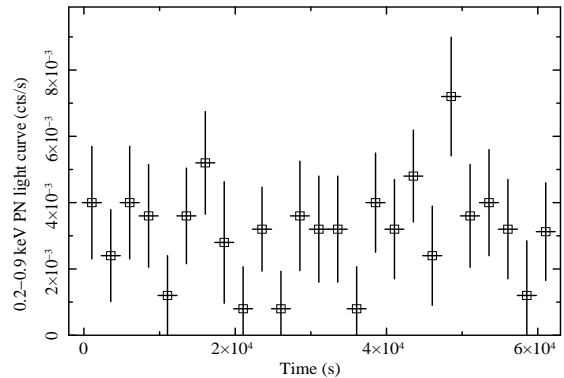}
\caption{The 0.2--0.9~keV pn light curve from the 2008 {\it
    XMM--Newton} observation of PHL~1092 with time bins of 2.5~ks.}
\label{lccomp}
\end{center}
\end{figure}

In Fig.~\ref{lccomp}  we show the EPIC pn lightcurve of PHL~1092 in
the 0.2--0.9~keV band. No short--timescale variability is
significantly detected in the whole $\sim$~60~ks light curve, while
variability of 50 per cent or more is normally observed on
similar timescales (see e.g. Brandt et al. 1999; Gallo et al. 2004).

The lack of X--ray variability could in principle indicate that a
different spectral component is responsible for the 2008 spectrum. It
is for instance possible that soft X--rays are associated with a
large--scale scattered/reprocessed component, overwhelmed by the
nuclear continuum emission in normal flux states. A rich spectrum of
soft X--ray emission lines is indeed observed in obscured Seyfert~2
(see e.g. Guainazzi \& Bianchi 2007) and in unobscured Seyfert~1
galaxies when observed at very low X--ray flux levels (e.g. NGC 4051,
Pounds et al. 2004, Ponti et al. 2006; Mrk 335, Grupe et al. 2008a;
Longinotti et al. 2008). This gas is likely associated with the Narrow
Line Region (Bianchi, Guainazzi \& Chiaberge 2006) thus responding to
continuum variation on very long timescales, and it typically
represents a few per cent of the soft 0.5--2~keV nuclear flux (Bianchi
\& Guainazzi 2007). In PHL~1092, the soft X--ray flux (or luminosity)
observed in 2008 is about 0.7 per cent of the historical, which means
that a scattering/reprocessing interpretation is viable provided that
the average flux of PHL~1092 in the past few hundreds years (assuming
a size of few hundreds parsecs for the reprocessing medium) was a
factor of a few smaller than that we observed over the past 30 years.

It should however be stressed that in all cases for which sufficiently
high quality data can be collected, accreting black holes satisfy the
so--called linear rms--flux relationship on all timescales (Uttley \&
McHardy 2001). This means that sources in low flux states are much
less likely to exhibit large amplitude variability than at higher flux
levels. Thus the apparent lack of X--ray variability is not
necessarily related to some physical difference in the X--ray
emission, but could simply be an extension to low flux levels of the
variability properties of the source.

\subsection{Simultaneous UV and quasi--simultaneous optical data}

All three {\it XMM--Newton} observations provide data from the OM. In
particular, exposures are available in the UVW2 filter (effective
wavelength of $\sim$1480~\AA~in the rest--frame). The OM data for the
2000 and 2003 observations were analysed by Gallo et al. (2004) who
report an average UVW2 flux of $4.43 \pm 0.05\times
10^{-15}$erg~s$^{-1}$~cm$^{-2}$~\AA$^{-1}$~in 2000 and of $\sim
3.79\pm 0.05\times 10^{-15}$erg~s$^{-1}$~cm$^{-2}$~\AA$^{-1}$~in
2003. Moreover, Gallo et al. (2004) also claim tentative evidence for
UV short--timescale variability of the order of few per cent,
especially during the 2003 observation. During our deep minimum 2008
observation, the UV flux is $(3.79\pm 0.05)\times
10^{-15}$erg~s$^{-1}$~cm$^{-2}$~\AA$^{-1}$~, i.e. consistent with the
2003 flux. The UV light curve (not shown here, 20 exposures of
$\sim$~2700~s each in UVW2) is consistent with a constant ($\chi^2_\nu
=0.8$) as is the X--ray lightcurve (see Fig.~\ref{lccomp}).
\begin{figure}
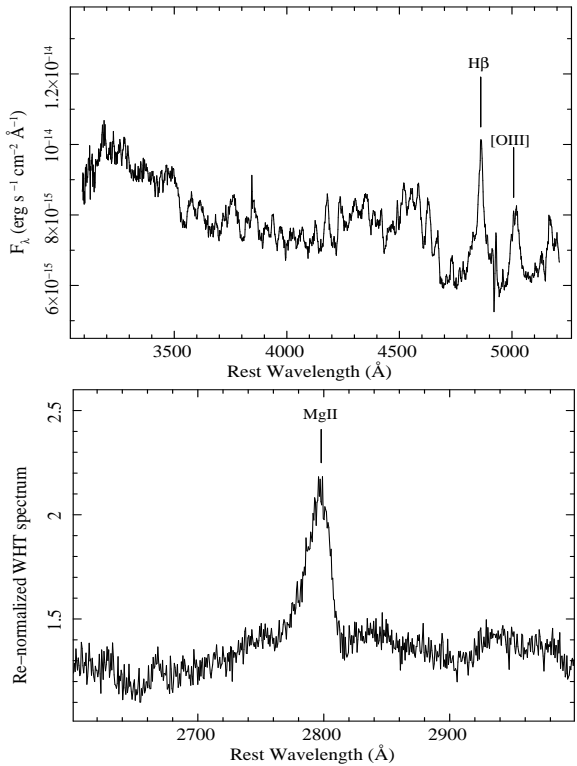

\begin{center}
\includegraphics[width=0.28\textwidth,height=0.42\textwidth,angle=-90]{het2.ps}
{\vspace{0.1cm}}
\includegraphics[width=0.28\textwidth,height=0.42\textwidth,angle=-90]{wht2.ps}
\caption{The quasi--simultaneous 2008 HET--LRS spectrum (top, no
  absolute flux calibration). We only identify the
  H$\beta$$\lambda$4861\AA~and [O~\textsc{iii}]$\lambda$5007\AA~lines
  as a reference (both affected by blended Fe~\textsc{ii}
  lines). Shorter wavelength lines mostly belong to Fe~\textsc{i} and
  Fe~\textsc{ii} optical/UV multiplets (see e.g. Bergeron \& Kunth
  (1980). In the bottom panel, the
  Mg~\textsc{ii}$\lambda$2798\AA~portion of the arbitrarily
  re--normalized WHT--ISIS spectrum is shown.}
\end{center}
\label{optical}
\end{figure}

Our UV flux measurement implies that the PHL~1092 spectral energy
distribution (SED) is now characterized by a much steeper optical to
X--ray spectral slope $\alpha_{\rm{ox}}$, defined as the hypothetical
power law slope between the 2500~\AA~ and 2~keV rest--frame flux
densities, than in 2003. A $\alpha^{(\rm{03})}_{\rm{ox}} \simeq -
1.56$ slope from the simultaneous 2003 UV and X--ray data was computed
by Gallo (2006). By applying the well known relationship between
$\alpha_{\rm{ox}}$ and the 2500~\AA~luminosity $l_{\rm{2500}}$
(e.g. Steffen et al. 2006; see also Vignali, Brandt \& Schneider 2003,
Strateva et al 2005; Gibson, Brandt \& Schneider 2008), we infer that
PHL~1092 should have $\alpha_{\rm{ox}} \simeq - 1.48$. The 2003 value
($-$1.56) is consistent with the spread in the relationship for
objects with $\log l_{\rm{2500}} \simeq 30-31$ (Steffen et al. 2006),
while in 2008 we measure $\alpha^{(\rm{08})}_{\rm{ox}} \simeq -2.44$,
well outside the $\alpha_{\rm{ox}}-l_{\rm{2500}}$ spread.

Following Gibson et al. (2008) we define $\Delta \alpha_{\rm{ox}}$ as
the difference between the observed $\alpha_{\rm{ox}}$ and the one that
can be predicted from the optical luminosity. While $\Delta^{\rm{03}}
\alpha_{\rm{ox}}= -0.08$ is quite typical for optically--selected QSO,
the 2008 value is $\Delta^{\rm{08}} \alpha_{\rm{ox}}= -0.96$, so
extreme that much less than 1 per cent of optically--selected QSO with
no Broad Absorption Line (BAL) systems are expected to share with
PHL~1092 such X--ray weakness (Gibson et al. 2008).

Quasi--simultaneous optical spectra were also obtained at the HET
(Low--Resolution Spectrograph, LRS) and WHT (Intermediate Dispersion
Spectrograph and Imaging System, ISIS), 10 and 25 days respectively
after the {\it XMM--Newton} observation. It is not the purpose of this
Letter to present detailed optical spectroscopy, but only to check
whether, given the huge X--ray flux variation, a corresponding
dramatic change in the optical properties of the source has
occurred. The HET spectrum ($\sim$3100--5250\AA~in the rest--frame) is
shown in Fig~6 (top). It is dominated by optical and UV Fe~\textsc{ii}
lines with relatively weak H$\beta$ and [O~\textsc{iii}], and does not
show any striking deviation from the one presented by Bergeron \&
Kunth (1980). The same is true for the WHT spectrum which extends down
to Mg~\textsc{ii} and up to $\sim$6500\AA~(rest--frame) the observed
wavelength range. We do not have coverage of the C~\textsc{iv} region
which would be interesting to search for the appearance of a BAL
signature in PHL~1092 (Brandt, Laor \& Wills 2000). We show however
the Mg~\textsc{ii}$\lambda$2798\AA~ region, which does not show the
developing of any low--ionization BAL (Fig.~6, bottom). We conclude
that no dramatic change in the optical properties of the source has
shown up on timescales of tens of days after the observed extreme
X--ray flux drop of PHL~1092.

\section{Discussion}

The major result of our {\it XMM--Newton} observation of PHL~1092 is
the discovery of a dramatic X--ray flux drop with respect to any
previous observation, together with a persistent constant
UV flux. This produces a very significant steepening of 
$\alpha_{\rm{ox}}$ from $-1.56$ to $-2.44$ in just
$\sim$~4.5 years. The X--ray weakness during 2008 is such that the
$\alpha_{\rm{ox}}$ deviation from a normal optically--selected QSO
with the same 2500~\AA~luminosity is one of the largest observed so
far for non--BAL QSO ($\Delta \alpha_{\rm{ox}} = -0.96$). 

A similar case in terms of extreme X--ray weakness is that of PHL~1811
($\Delta \alpha_{\rm{ox}} = -0.70$; Leighly et al. 2007a,b; see also
LBQS~0102-2713 for which Boller et al. 2009 report $\Delta
\alpha_{\rm{ox}} \simeq -0.50$). Both PHL~1092 and PHL~1811 represent
the high--luminosity end of the NLS1 population (with PHL~1811 being
about 5 times more optically luminous than PHL~1092). Moreover, they
both have strong {Fe\,\textsc{ii}} emission, relatively weak
[~{O\,\textsc{iii}}~] and unusually blue far--UV spectra, while
lacking strong Ly$\alpha$ and {C\,\textsc{iv}} lines (Leighly et
al. 2007a,b). However, PHL~1811 has always been observed to be X--ray
weak over the past $\sim$20 years (in 7 different observations) while,
for a comparable number of X--ray observations and monitoring period,
PHL~1092 has been observed to be extremely X--ray weak only once. Our
2008 observation of PHL~1092 implies that transitions from relatively
standard QSO states (in terms of $\alpha_{\rm{ox}}$) to extreme X--ray
weak ones are possible on relatively short timescales (few
years). This in turns means that outliers in the
$\alpha_{\rm{ox}}-l_{\rm{2500}}$ relation may be due to transient
extreme X--ray variability phenomena and that they may change their
status if sufficiently long monitoring is performed.

Despite the factor of $\sim 200$ X--ray flux drop of PHL~1092, the
persistent soft spectral shape, combined with the non--detection of
(simple) intrinsic X--ray absorption, may suggest that an absorption
event is unlikely. However, As discussed in Section~3.2, the nature
of the soft X--ray emission in the 2008 observation of PHL~1092 is
uncertain (either direct nuclear emission or large--scale
scattered/reprocessed emission). If X--rays are scattered, the nuclear
emission could be absorbed by e.g. Compton--thick matter. If so, the
persistent UV flux in PHL~1092 means that the absorber would have to
completely stop X--rays (e.g. being Compton--thick) while being
transparent in the UV regime, which seems highly unlikely given that
the the UV and X--ray emission regions (accretion disc and X--ray
corona, respectively in the standard view) most likely share similar
locations. Even by considering unusual absorber size and covering
factor or non--standard physical properties (ionization state, grain
size distribution) it seems difficult that UV and optical would be
unaffected. One possibility would be that the UV are also scattered so
that they are spatially separated from the X--ray emitting
region. However, tentative evidence for short timescale UV variability
during the 2003 (Gallo et al. 2004) argues against such a hypothesis.

A partial covering scenario has been invoked to account for the X--ray
weakness of some other NLS1 galaxies (e.g. Mrk~335, Grupe et
al. 2008a; WPVS~007, Grupe, Leighly, Komossa 2008b). However, UV
variability was always observed, as opposed to the present
case. Another possible scenario is one in which the direct nuclear
emission is suppressed by light--bending effects near the black hole
(Miniutti \& Fabian 2004; Fabian et al. 2004). Light bending can be
responsible for a maximum X--ray variability of a factor $\sim 20$,
and the remaining factor $\sim 10$ should have to be accounted for by
intrinsic variability. This is more acceptable especially for NLS1
sources, but we consider somewhat unlikely that both effects conspire
to act simultaneously and in the same direction.

Generally speaking, the UV to X--ray spectral energy distribution of
AGN is interpreted in terms of accretion disc emission (peaking in the
UV) and inverse Compton by one or more population of hot electrons
(the so--called X--ray corona) upscattering the soft UV seed photons
up to X--ray energies. Since the UV flux remains constant between the
2003 and 2008 observations, if we exclude an absorption origin for the
sudden X--ray weakness of PHL~1092, we must conclude that the ability
of the X--ray corona to upscatter the optical/UV seed photons through
inverse Compton was very significantly reduced during the 2008
observation (see Grupe et al. 2000 for a similar explanation in the
case of the radio--loud NLS1 galaxy RX~J0134.2--4258). In order to get
insights on the plausibility of genuine X--ray coronal variability in
PHL~1092, it may be useful to compare the behaviour of PHL~1092 with
that of Galactic black hole binaries. The much smaller size of these
systems allows one to probe timescales that are not accessible for
AGN, thus providing potential clues to AGN studies. By making use of
the analogy between the AGN and Galactic black hole SED, the question
here would be to consider whether Galactic sources can exhibit large
amplitude variability events in the X--ray corona component (power
law) while the soft X--ray disc component is stable.

In a recent work Sobolewska, Gierlinski \& Siemiginowska (2009)
introduced a $\alpha^{'}_{\rm{GBH}}$ parameter defined as the
equivalent of $\alpha_{\rm{ox}}$ for Galactic black hole binaries,
i.e. the quantity measuring the spectral slope between the disc-- and
corona--dominated components. They show that $\Delta
\alpha^{'}_{\rm{GBH}}$ of the order of unity can be reached by a
single source at a given disc luminosity during a given outburst. The
observed timescale for such variations is of the order of
days. However, to the best of our knowledge, no firm constraints can
be placed on the shortest possible timescales, making it possible
(though most probably very rare) that variability of this amplitude
could be seen in AGN on timescales of years. A small mass for the
black hole in PHL~1092 would help to reduce its variability
timescale. In fact, while its optical luminosity would suggest a black
hole mass of the order of $1-2\times 10^8~M_\odot$ (Dasgupta, Rao \&
Dewangan 2004), X--ray variability suggests instead a much smaller
mass of $\sim 10^6~M_\odot$ (Bian \& Zhao 2003; Nikolajuk, Czerny \&
Gurynowicz 2009). Although extrinsic explanations such as absorption
or light bending cannot be firmly ruled out, we conclude that we have
likely observed a rare and extreme variability event associated with a
transient dramatic weakening or disruption of the X--ray corona in
PHL~1092. Future X--ray and optical/UV monitoring of the source is
mandatory to reveal the nature of the extreme event we have reported
here.

\section*{Acknowledgements}

Based on observations obtained with XMM-Newton, an ESA science mission
with instruments and contributions directly funded by ESA Member
States and NASA. GM thanks the Ministerio de Ciencia e Innovaci\'on
and CSIC for support through a Ram\'on y Cajal contract.  ACF thanks
the Royal Society for support. WNB thanks NASA grant NNX08AW87G. We
also thank Brendan Miller for helping with the HET data and John
Southworth for taking the WHT spectrum of PHL~1092. We also thank
Richard Saxton for the {\it XMM--Newton} Slew upper limit.


\begin{thebibliography}{}

\bibitem[Bergeron \& Kunth(1980)]{1980A&A....85L..11B} Bergeron J.,
  \& Kunth D.\ 1980, \aap, 85, L11

\bibitem[Bergeron \& Kunth(1984)]{1984MNRAS.207..263B} Bergeron J.,
  \& Kunth D.\ 1984, \mnras, 207, 263

\bibitem[Bian 
\& Zhao(2003)]{2003ApJ...591..733B} Bian W., Zhao Y.\ 2003, \apj, 591, 733

\bibitem[Bianchi et 
al.(2006)]{2006A&A...448..499B} Bianchi S., Guainazzi M., Chiaberge M., 2006, \aap, 448, 499

\bibitem[Bianchi 
\& Guainazzi(2007)]{2007AIPC..924..822B} Bianchi S., Guainazzi M., 2007, Proc. of the conference ``The Multicolored Landscape of Compact Objects and Their Explosive Origins'', AIP Conf. Proc., 924, 822

\bibitem[Boller et 
al.(1996)]{1996A&A...305...53B} Boller T., Brandt W.~N.,  Fink H.\ 1996, \aap, 305, 53

\bibitem[Boller et al.(1997)]{1997MNRAS.289..393B} Boller T., Brandt 
W.~N., Fabian A.~C., Fink H.~H.\ 1997, \mnras, 289, 39

\bibitem[Boller et al.(1996)]{boller09} Boller T., Linguri K., Heftrich T., Weigand M., 2009, submitted to ApJ

\bibitem[Brandt et al.(1997)]{1997MNRAS.285L..25B} Brandt W.~N., Mathur 
S., \& Elvis M.\ 1997, \mnras, 285, L25

\bibitem[\protect\citeauthoryear{Brandt et al. }{1999}]{brandt99}
  Brandt W.N., Boller Th., Fabian A.C., Ruszkowsi M., 1999, MNRAS,
  303, L53

\bibitem[Brandt et al.(2000)]{2000ApJ...528..637B} Brandt W.~N., Laor A., 
Wills B.~J.\ 2000, \apj, 528, 637

\bibitem[Dasgupta et al.(2004)]{2004ApJ...614..626D} Dasgupta S., Rao 
A.~R., Dewangan G.~C.\ 2004, \apj, 614, 626

\bibitem[Fabian et al.(2004)]{2004MNRAS.353.1071F} Fabian A.~C., Miniutti 
G., Gallo L., Boller T., Tanaka Y., Vaughan S., Ross R.~R.\ 2004, \mnras, 353, 1071

\bibitem[Forster \& Halpern(1996)]{1996ApJ...468..565F} Forster K.,
  Halpern J.~P., 1996, \apj, 468, 565

\bibitem[\protect\citeauthoryear{Gallo et al. }{2004}]{gallo04}
  Gallo L.C., Boller Th., Brandt W.N., Fabian A.C., Grupe D., 2004,
  MNRAS, 352, 744

\bibitem[Gallo(2006)]{2006MNRAS.368..479G} Gallo L.~C.\ 2006, \mnras, 368, 
479 

\bibitem[Gibson et al.(2008)]{2008ApJ...685..773G} Gibson R.~R., Brandt 
W.~N., Schneider D.~P.\ 2008, \apj, 685, 773 

\bibitem[\protect\citeauthoryear{Goodrich }{1989}]{good89}
Goodrich R.W., 1989, \apj, 342, 224


\bibitem[Grupe et al.(2000)]{2000A&A...356...11G} Grupe D., Leighly
  K.~M., Thomas H.-C., Laurent-Muehleisen S.~A., 2000, \aap, 356, 11

\bibitem[Grupe et al.(2008)]{2008ApJ...681..982G} Grupe D., Komossa S., 
Gallo L.~C., Fabian A.~C., Larsson J., Pradhan A.~K., Xu D., Miniutti G.\ 2008a, \apj, 681, 982

\bibitem[Grupe et al.(2008)]{2008AJ....136.2343G} Grupe D., Leighly 
K.~M., Komossa, S.\ 2008b, \aj, 136, 2343 


\bibitem[Guainazzi 
\& Bianchi(2007)]{2007MNRAS.374.1290G} Guainazzi M.,  Bianchi S.\ 2007, \mnras, 374, 1290 

\bibitem[Kwan et al.(1995)]{1995ApJ...440..628K} Kwan J., Cheng F.-Z., 
Fang L.-Z., Zheng W., Ge J.\ 1995, \apj, 440, 628

\bibitem[Leighly(1999)]{1999ApJS..125..297L} Leighly K.~M.\ 1999a, \apjs, 
125, 297

\bibitem[Leighly(1999)]{1999ApJS..125..317L} Leighly K.~M.\ 1999b, \apjs, 
125, 317

\bibitem[Leighly et al.(2007)]{2007ApJ...663..103L} Leighly K.~M., 
Halpern J.~P., Jenkins E.~B., Grupe D., Choi J., Prescott K.~B.\ 2007a, \apj, 663, 103

\bibitem[Leighly et al.(2007)]{2007ApJS..173....1L} Leighly K.~M., 
Halpern J.~P., Jenkins E.~B., Casebeer D.\ 2007b, \apjs, 173, 1

\bibitem[Longinotti et 
al.(2008)]{2008A&A...484..311L} Longinotti A.~L., Nucita A., Santos-Lleo M., Guainazzi M.\ 2008, \aap, 484, 311

\bibitem[Miniutti 
\& Fabian(2004)]{2004MNRAS.349.1435M} Miniutti G., Fabian A.~C.\ 2004, \mnras, 349, 1435

\bibitem[Murphy et al.(1996)]{nh} 
Murphy E.M., Lockman R.J., Laor A., Elvis M., 1996, ApJS, 105, 369

\bibitem[Nikolajuk et al.(2009)]{2009arXiv0901.1442N} Nikolajuk M., 
Czerny B., Gurynowicz P.\ 2009, MNRAS in press (arXiv:0901.1442)

\bibitem[Osterbrock \& Pogge(1985)]{1985ApJ...297..166O} Osterbrock
  D.~E.,  Pogge R.~W.\ 1985, \apj, 297, 166

\bibitem[Ponti et al.(2006)]{2006MNRAS.368..903P} Ponti G., Miniutti G., 
Cappi M., Maraschi L., Fabian A.~C., 
\& Iwasawa K.\ 2006, \mnras, 368, 903

\bibitem[Pounds et al.(2004)]{2004MNRAS.350...10P} Pounds K.~A., Reeves 
J.~N., King A.~R.,  Page K.~L.\ 2004, \mnras, 350, 10 

\bibitem[Sobolewska et al.(2009)]{2009arXiv0901.2150S} Sobolewska M.~A., 
Gierlinski M., Siemiginowska A., 2009, MNRAS, 394, 1640

\bibitem[Steffen et al.(2006)]{2006AJ....131.2826S} Steffen A.~T., 
Strateva I., Brandt W.~N., Alexander D.~M., Koekemoer A.~M., Lehmer 
B.~D., Schneider D.~P., Vignali C.\ 2006, \aj, 131, 2826

\bibitem[Strateva et al.(2005)]{2005AJ....130..387S} Strateva I.~V., 
Brandt W.~N., Schneider D.~P., Vanden Berk D.~G., 
Vignali C.\ 2005, \aj, 130, 387 

\bibitem[Uttley \& McHardy 2001]{rmsflux} 
Uttley P., McHardy I.M., 2001, \mnras, 323, L26

\bibitem[V{\'e}ron-Cetty et al.(2001)]{2001A&A...372..730V}
  V{\'e}ron-Cetty M.-P., V{\'e}ron P.,  Gon{\c c}alves
  A.~C.\ 2001, \aap, 372, 730

\bibitem[Vignali et al.(2003)]{2003AJ....125..433V} Vignali C., Brandt 
W.~N., Schneider D.~P.\ 2003, \aj, 125, 433 

\bibitem[Wilkes et al.(1994)]{1994ApJS...92...53W} Wilkes B.~J., 
Tananbaum H., Worrall D.~M., Avni Y., Oey M.~S., 
Flanagan J., 1994, \apjs, 92, 53

\end{thebibliography}
\end{document}